# Spin-1/2 kagome compounds: volborthite vs herbertsmithite

Z Hiroi[1], H Yoshida, Y Okamoto and M Takigawa
Institute for Solid State Physics, University of Tokyo,
Kashiwa, Chiba 277-8581, Japan

E-mail: hiroi@issp.u-tokyo.ac.jp

**Abstract.** Two kagome compounds, volborthite $Cu_3V_2O_7(OH)_2 \cdot 2H_2O$ and herbertsmithite $ZnCu_3(OH)_6Cl_2$, are compared in order to derive information about the intrinsic properties of the spin-1/2 kagome antiferromagnet. Volborthite shows a broad maximum at $T \sim J/4$ and the approach at $T = 0$ to a large finite value of the bulk magnetic susceptibility $\chi_{bulk}$ as well as the local susceptibility $\chi_{local}$ from NMR measurements. These must be intrinsic properties for the spin-1/2 kagome antiferromagnet, as similar behavior has also been reported in $\chi_{local}$ for herbertsmithite [Olariu A *et al*. 2008 *Phys. Rev. Lett.*, **100** 087202]. Impurity effects that may significantly influence the bulk properties are discussed.

## 1. Spin-1/2 kagome antiferromagnet

A kagome lattice is one of the typical playgrounds for frustration physics. Extensive theoretical and experimental study has been carried out to understand the property of antiferromagnetically coupled spins on the kagome lattice. However, the ground state (GS) of the most attractive kagome antiferromagnet (KAFM) with $S$-1/2 Heisenberg spins seems to be beyond our understanding. This is because of the inherent difficulty in carrying out theoretical treatments of the effects of frustration [1] as well as the lack of ideal model compounds to study experimentally [2].

The theoretically expected GS of the $S$-1/2 KAFM is a sort of singlet state with an energy gap in the spin excitation spectrum [1, 3-5]. Since the predicted magnitude of the gap $\Delta$ is small, $J/4$ or $J/20$, one assumes that there exists an unusual state covered by extended singlet pairs instead of local singlets that are often observed for various compounds in which a bond alternation exists or is induced as a result of spin-lattice coupling. It is known that the size of the singlet pairs or the correlation length $\xi$ is inversely proportional to $\Delta$; the smaller the $\Delta$, the larger the $\xi$. Determining the magnitude of the gap should be crucial to understanding the GS of the $S$-1/2 KAFM, though it seems difficult to estimate precisely from theoretical studies.

On the other hand, it is always a challenge for materials scientists to find an ideal model compound. Real compounds inevitably suffer from more or less disorder arising from defects in a crystal and unwanted anisotropy or the three dimensionality of interactions. Here we focus on the two candidate compounds studied recently and believed to approximate to $S$-1/2 KAFMs. Both are copper minerals with spin 1/2 carried by a $Cu^{2+}$ ion. One is volborthite $Cu_3V_2O_7(OH)_2 \cdot 2H_2O$ [6-9] which possesses a kagome layer comprising edge-sharing octahedra, as depicted in Fig. 1(a). Since it crystallizes in a monoclinic structure [6], there is an anisotropy in the magnetic interactions between nearest-neighbour $Cu^{2+}$ spins in the plane. The distances between pairs of Cu atoms are 3.031 (Cu1-Cu2) and 2.937 Å

---
[1] To whom any correspondence should be addressed.

(Cu2-Cu2). The Cu2 ion is located in an octahedron made of 4 O and 2 OH ions that is elongated horizontally in Fig. 1(a), while the octahedron of the Cu1 ion is deformed in the opposite sense. Thus, it is reasonable to assume that an unpaired electron is in the $d_{z^2}$ orbital at Cu1, but in the $d_{x^2-y^2}$ orbital of Cu2, as schematically drawn in the inset of Fig. 1(a). As a result, moderately strong antiferromagnetic superexchange couplings are expected through bridging oxide ions with large Cu-O-Cu angles; these are 105.6° and 82.7° for $J_1$ between Cu1 and Cu2 spins, and 101.1° and 91.5° for $J_2$ between two Cu2 spins [6]. Although it is difficult to predict the magnitude of magnetic couplings, the disparity may not be so large, because of this orbital arrangement. In fact, recent theoretical calculations on the magnetic susceptibility and specific heat of volborthite suggested that the lattice retains a great deal of frustration [10] and that the anisotropy can be less than 20% [11]. The average coupling $J_{av} = (2J_1 + J_2) / 3$ was estimated to be 84 K in our previous study [7]. On the one hand, an anisotropic kagome model has been studied theoretically and shows a rich phase diagram with ferrimagnetic, incommensurate and decoupled chain phases [12].

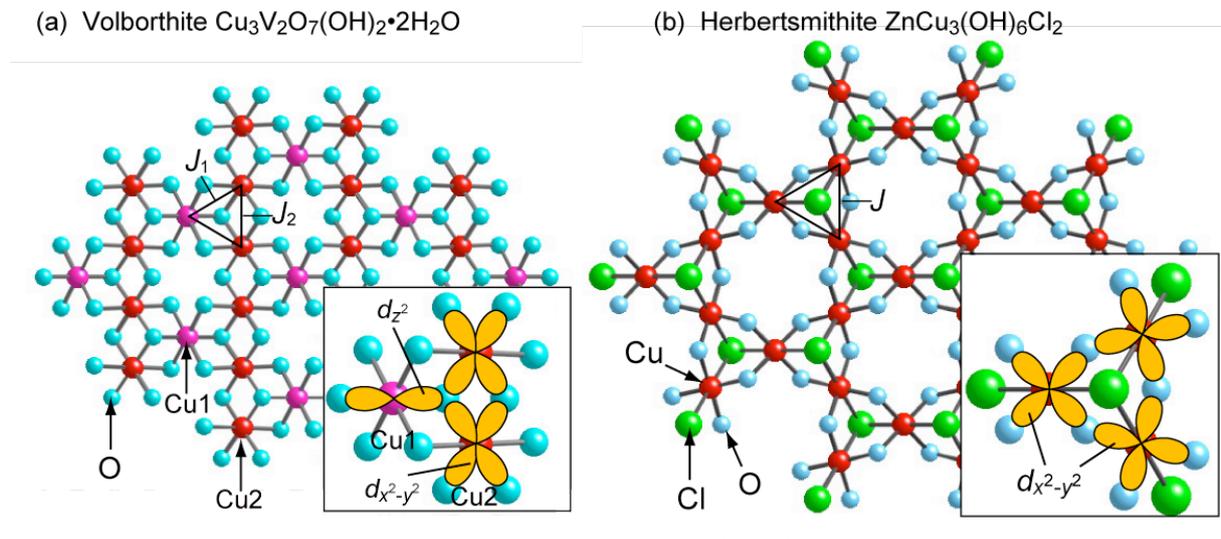

**Figure 1.** Kagome lattices of volborthite (a) and herbertsmithite (b). The two drawings are on the same scale. The inset on each drawing expands a triangle of Cu ions to show a possible arrangement of 3$d$ orbitals carrying unpaired electrons.

The other compound is herbertsmithite $ZnCu_3(OH)_6Cl_2$ which was claimed to be a structurally perfect KAFM [13]. In fact, it crystallizes in a rhombohedral structure, as depicted in Fig. 1(b), comprising an equilateral triangle made of $Cu^{2+}$ ions [14]. The Cu-Cu distance is 3.414 Å, more than 10% larger than those in volborthite. Magnetic coupling in the triangle should be the same through a nearest-neighbour superexchange $J$ via a Cu-O-Cu path based on the $d_{x^2-y^2}$ orbitals arranged symmetrically along the threefold axis. The magnitude of $J$ was estimated to be 170 ~ 190 K [15-17], more than double the $J_{av}$ of volborthite. This is due to the larger bond angle of 119° [18]. Although the compound appears to be perfect, its Achilles heel is a mutual exchange between $Cu^{2+}$ and nonmagnetic $Zn^{2+}$ ions [19-21]. It was reported that 6 - 10% of the Cu site in the kagome plane is replaced by Zn, which means that 18 - 30% of the Zn sites are occupied by Cu ions. This may be caused by the similarity in the ionic radii of $Cu^{2+}$ and $Zn^{2+}$ and also the two ions having the same valence state. The associated disorder effects in the kagome plane must seriously disturb the GS. Moreover, the almost free Cu spins at the Zn site mask the intrinsic properties in bulk measurements: a superexchange between two neighboring Cu spins in the Cu and Zn sites is expected to be relatively small because of the particular Cu-O-Zn bond angle of 96.9° [18]. Such a chemical substitution is not the case for volborthite, because a mutual exchange between aliovalent $Cu^{2+}$ and $V^{5+}$ ions is unfavourable in terms of ionic radius and Madelung energy.

Neither compound exhibits long-range order down to 50 mK so they have been presumed to be in a spin liquid state [8, 22]. In this paper, we compare the uniform and local magnetic susceptibilities of the two compounds and try to clarify the similarity and differences between them. Throughout the comparison, we discuss a possible GS for the ideal spin-1/2 KAFM.

## 2. Experimental
Polycrystalline samples of volborthite and herbertsmithite were prepared as previously reported for each compound [7, 13, 14]. Magnetic susceptibility was measured using a commercial SQUID magnetometer in magnetic fields of 1 and 10 kOe between 2 and 400 K.

## 3. Magnetic susceptibility
Figure 2 compares the magnetic susceptibility $\chi$ for the two compounds. Marked features are a broad maximum observed at around $T_{peak}$ = 22 K for volborthite and a large Curie tail at low temperatures for herbertsmithite. The former may be associated with the development of antiferromagnetic short-range-order (SRO) [7], and the latter is apparently due to free spins mostly on the Zn site. Both compounds exhibit Curie-Weiss (CW) behaviour at high temperatures, as clearly evidenced in the inverse-$\chi$ plot shown in the inset. The Weiss temperature $\Theta_W$ and the Lande $g$ factor are $\Theta_W$ = -115 K and $g$ = 2.26 for volborthite and $\Theta_W$ = -241 K and $g$ = 2.23 for herbertsmithite. The latter value is slightly smaller than the value $\Theta_W \sim$ -300 K reported previously [13, 15]. The difference in the magnitude at high temperatures comes from the difference in the Weiss temperature and thus the antiferromagnetic interaction.

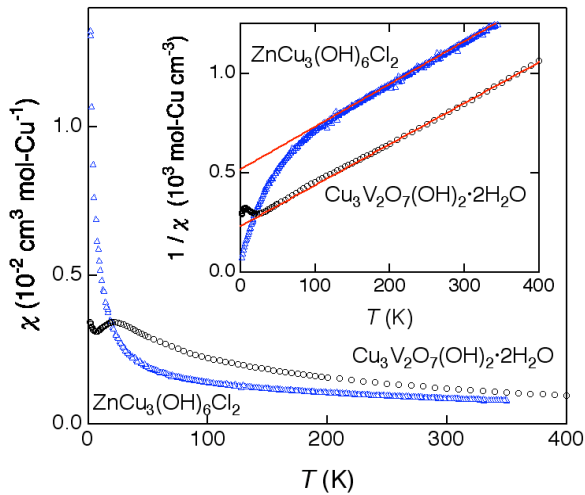

**Figure 2.** Magnetic susceptibility $\chi$ per mole of Cu for volborthite and herbertsmithite measured at $H$ = 10 kOe with temperature increasing. The inset shows the same data plotted as a function of $1/\chi$. The line through each data set denotes a Curie-Weiss fit.

Various fittings to the $\chi$ data have been carried out, as shown in Fig. 3. The $J_{av}$ of volborthite is determined to be $J_{av} / k_B$ = 86 K by fitting the data above $T$ = 150 K to the calculations by high-temperature series expansions (HTSE) for the kagome lattice based on the spin Hamiltonian $J \sum S_i \cdot S_j$ [3]. A Bonner-Fisher (BF) curve assuming $J_{av} / k_B$ = 136 K cannot reproduce the observed steep increase at low temperatures as well as the sharper peak at 22 K, which means that the Heisenberg chain model is far from the reality. Taking into account couplings between chains would improve the result and may lessen the anisotropy on the kagome lattice, as discussed before [11]. It is important to note that the value of $\chi$ can remain large and finite at $T$ = 0, implying the absence of a spin gap above the GS. After subtracting a small upturn at low temperature that is ascribed to 0.5% free spins, the remaining value becomes 2.7 × 10$^{-3}$ cm$^3$ / mol-Cu. If there was a gap of $J/4$ as predicted by the finite-cluster calculations [3], $\chi$ should rapidly decrease after a maximum at $J/6 \sim 14$ K, as shown in Fig. 3(a). Thus, we can safely exclude the presence of such a large gap. Very recently, we extended our measurements down to 60 mK on a higher-quality sample and found no downturn in $\chi$ [23]. Therefore, it is plausible to conclude that the GS of volborthite is gapless or, more precisely, that the spin gap is less than $J/1500$.

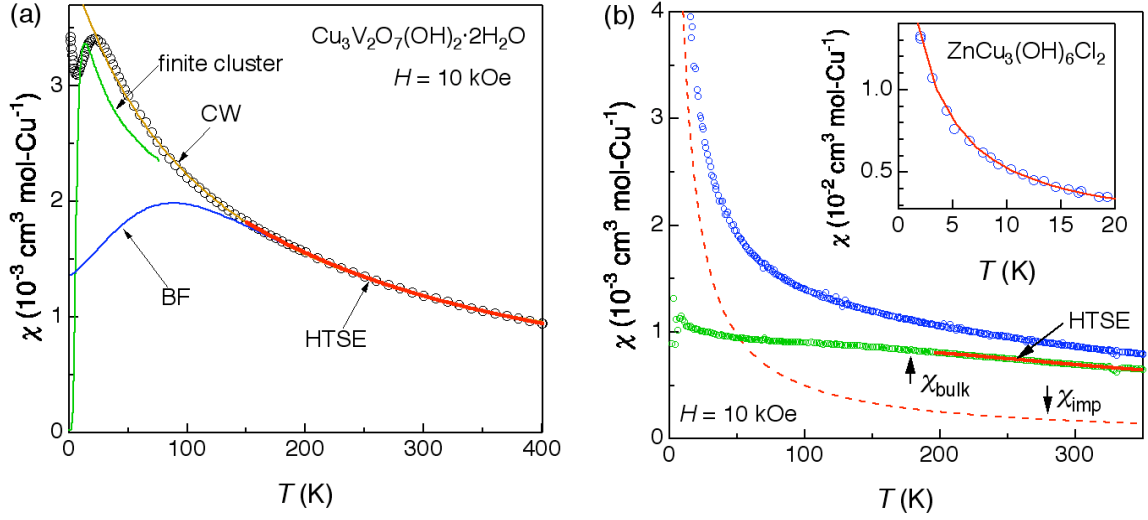

**Figure 3.** Various analyses on the magnetic susceptibilities of volborthite (a) and herbertsmithite (b). Shown in (a) are fits of the Curie-Weiss (CW) form, the Heisenberg chain model expressed by the Bonner-Fisher (BF) curve [24], and the uniform kagome lattice model by high-temperature series expansions (HTSE) [3]. The $T$ dependence expected for $\Delta = J/4$ by the finite cluster calculation is also plotted in (a). The inset of (b) expands the low-temperature part, where fitting to the form $\chi = \chi_{imp} + \chi_{bulk}$ is shown. The separate contributions are shown in the main panel below the data. The HTSE fits at high temperatures yield $J_{av}/k_B = 86$ K for volborthite and $J/k_B = 199$ K for herbertsmithite.

It is curious to know whether the broad maximum in $\chi$ observed for volborthite is a general feature for the spin-1/2 HAFM. In the case of herbertsmithite, the large Curie tail might have absorbed it into the background, even if it exists. Thus, it is reasonable to assume that $\chi = \chi_{imp} + \chi_{bulk}$, where $\chi_{imp} = C/(T - \Theta)$. We fitted the data between 2 and 17 K by assuming that $\chi_{bulk}$ is constant over that $T$ range to estimate the concentration of nearly free spins $x_{imp}$. The fitting yields $x_{imp} = 10.9(4)\%$ with $\chi_0 = 1.1(1) \times 10^{-3}$ cm$^3$ / mol-Cu and $\Theta = -2.2(1)$ K, which means that the actual compositions are Cu$_{0.9}$Zn$_{0.1}$ and Zn$_{0.7}$Cu$_{0.3}$ at the kagome Cu site and the Zn site, respectively, in good agreement with previous results from structural refinements [19, 20]. After subtraction of this large impurity contribution, the rest, $\chi_{bulk}$, is revealed to be flatter and does not clearly exhibit a broad peak as in volborthite. It is likely, however, that the ambiguity of estimating $\chi_{imp}$ at low temperatures has obscured a possible gradual decrease below ~50 K. By fitting $\chi_{bulk}$ between 200 and 350 K to the result of the HTSE, we obtain $J = 199(1)$ K assuming $g = 2.23$ from the CW fitting at high temperatures. This $J$ value is close to those reported previously [15-17]. Thus, the energy scale is more than twice as large in herbertsmithite than in volborthite.

In the case that the $\chi_{bulk}$ is influenced by impurity contributions, the NMR Knight shift $K$ often provides a good local probe to estimate $\chi_{local}$. In the case of volborthite, $\chi_{bulk}$ and $K$ from V NMR signals gave similar $T$ dependences that are plotted together in Fig. 4(a), taking into account the hyperfine coupling constant $A = 6.6$ kOe / $\mu_B$ in the relationship $K = A\chi / N_A$ [7]. Thus, the presence of a broad peak in the magnetic susceptibility is clearly evident, though $T_{peak}$ at the maximum seems to be slightly different between $\chi_{bulk}$ and $\chi_{local}$. Very recent $^{17}$O NMR experiments on herbertsmithite by Olariu *et al.* successfully determined the local susceptibility of the kagome lattice and found that $K$ decreases below 50 K and approaches approximately one-third of the maximum value at 50 K [21]. Their $\chi_{local}$ data is compared with our $\chi_{bulk}$ data in Fig. 4 using $A = 35$ kOe / $\mu_B$. Note that the $T_{peak}$ of herbertsmithite is considerably higher than that of volborthite, reflecting the larger $J$ value. Figure 4(b) shows an alternative plot after normalization using $J / k_B = 86$ (199) K and $g = 2.26$ (2.23) for volborthite (herbertsmithite). All the data except the $\chi_{bulk}$ of herbertsmithite merge into a universal curve at high temperatures, whereas their low-temperature behaviors seem different. However, we can say qualitatively that there is a broad maximum at around $T \sim J / 4$, indicating the development of SRO, and also that a large value remains at $T = 0$, supporting a gapless GS.

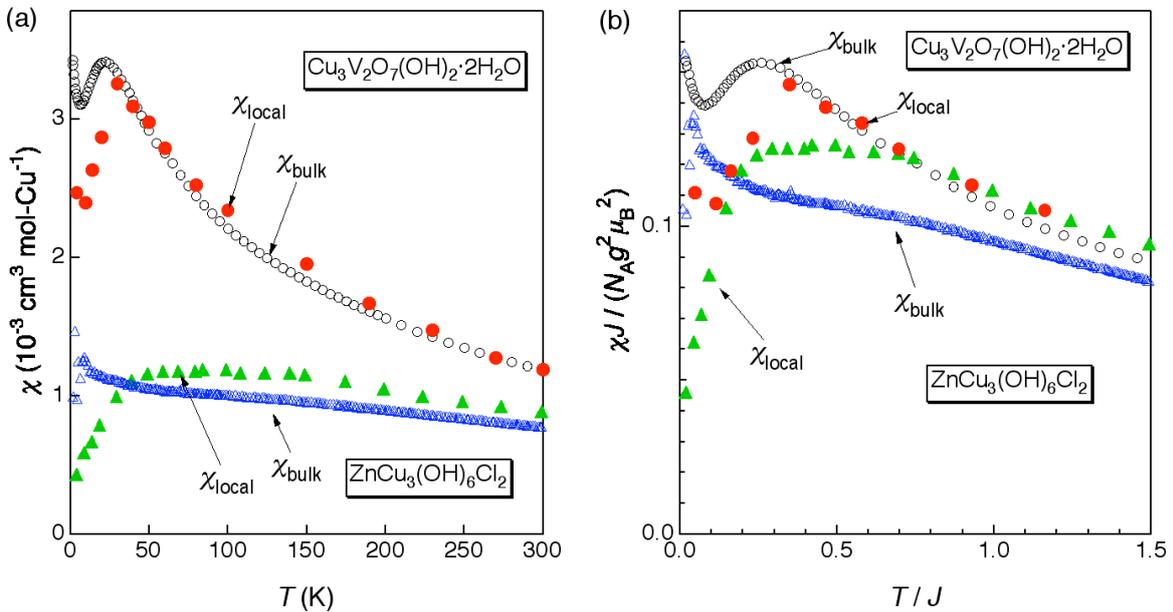

**Figure 4.** (a) Local magnetic susceptibility $\chi_{local}$ estimated from the Knight shift obtained by previous NMR measurements [7, 21]. The original Knight shift data were measured at $H \sim 80$ kOe for volborthite and $\sim 70$ kOe for herbertsmithite, while the magnetic susceptibility was obtained at $H = 10$ kOe. The $\chi_{bulk}$ data has been divided by 0.9 for comparison, taking into account the missing 10% spins from the kagome plane. (b) The same data plotted after normalization using $J / k_B = 86$ (199) K and $g = 2.26$ (2.23) for volborthite (herbertsmithite).

### 4. Impurity effects
Impurity effects are inevitable in real materials and may be crucial for understanding the true GS of the kagome compounds. Bert *et al.* found a spin-glass transition at 1.2 K for volborthite [25], as reproduced in Fig. 5. Their sample seems to contain more impurity spins than ours, judging from the larger Curie tail. Recently, we also observed a similar spin-glass transition in $\chi$ measurements down to 60 mK on our samples [23]. In the previous NMR study down to 1.7 K, we obtained a characteristic NMR spectrum that is composed of a sharp central peak and a broad hump, as reproduced in Fig. 6 [7]. Bert *et al.* also found a similar NMR spectrum and suggested that 20% of Cu spins are in SRO and 40% are frozen [25]. In contrast, we had ascribed the broad hump to field-induced paramagnetic moments, because its line width appeared to vanish at zero field, as reproduced in the inset of Fig. 6 [7]. It is known for spin-1/2 Heisenberg chains, such as $Sr_2CuO_3$, that local staggered magnetization is induced by the magnetic field near chain ends or defects, which gives a similar spectrum [26]. By analogy, we think that the broad hump comes from Cu spins that are located near and affected by a defect, as schematically illustrated in Fig. 7, and the sharp central peak is from the remaining intact spins that must carry the intrinsic information of the kagome lattice. It is considered that the spin glass transition observed at low temperature is related to moments induced by this defect. It is possible that an impurity-affected domain spreading over a distance $\xi$ from a defect grows with decreasing temperature and finally overlaps with nearby domains so as to be connected to each other percolatively, resulting in the spin glass transition.

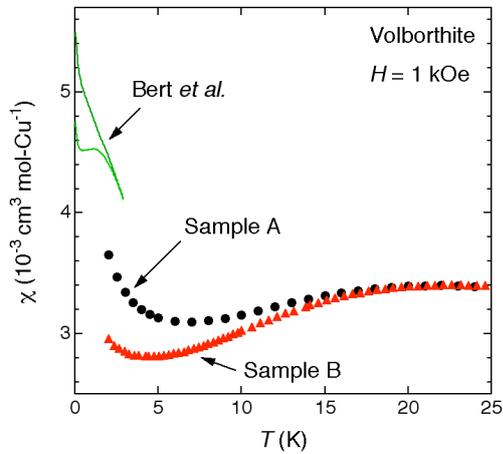

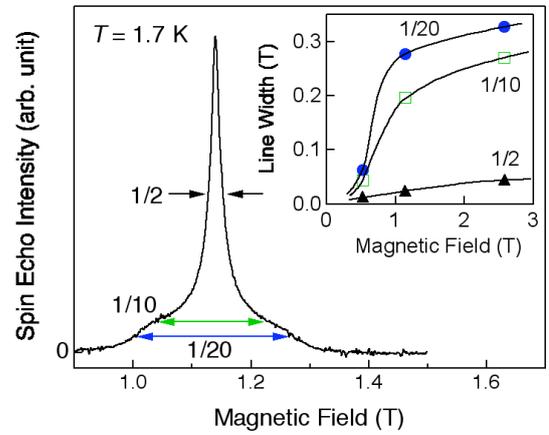

**Figure 5.** Comparison of the magnetic susceptibility of volborthite measured at $H = 1$ kOe; sample A was studied in the present paper and sample B was obtained by annealing in hydrothermal conditions [23]. The amounts of impurity spins are estimated from the Curie tails to be 0.5% and 0.07%, respectively. The $\chi$ data given by Bert et al. is also plotted [25], which exhibits a thermal hysteresis due to a spin-glass transition.

**Figure 6.** $^{51}$V powder NMR spectrum of volborthite measured at $T = 1.7$ K and $f = 12.8$ MHz [7]. The inset shows the field dependence of the full width at 1/2, 1/10 and 1/20 of the peak height.

It is therefore critically important to reduce the amount of impurity spins or defects as much as possible. Very recently, we have tried to improve the sample quality of volborthite and succeeded by annealing under hydrothermal conditions after the initial precipitation [23]. The x-ray diffraction peaks from the annealed sample become much sharper and the particle size approaches 100 μm. The value of $\chi$ measured on the new sample (sample B) is shown in Fig. 5, which exhibits a much smaller Curie tail and more pronounced decrease below $T_{peak}$. The estimation of $x_{imp}$ gives 0.07%, smaller by one order of magnitude compared with the present sample A. The identity of the defects in volborthite is not known, but may be associated with certain crystalline defects produced near the particle surface or at a stacking fault, as is often the case for such a layered compound. Various experiments are in progress on the clean sample to elucidate more clearly the nature of the GS of volborthite [23].

In strong contrast, it seems difficult to control the impurity level in herbertsmithite; mutual exchange between Cu and Zn atoms may occur, giving a high-entropy state at the preparation temperatures. We found that approximately 10% of Cu spins are missing from the kagome plane. Previous specific heat and NMR measurements found approximately 6% missing [20, 21]. It was suggested that about 20% of the remaining spins behave differently from the majority spins, because each defect directly affects four nearby Cu spins [21]. The situation for the 10% exchange is schematically depicted in Fig. 7, which clearly illustrates how imperfect the kagome plane of the herbertsmithite is. As observed in volborthite, even less than 1% of defects disturbs the surrounding spins over a large distance. One has to be careful to interpret experimental results from such a diluted kagome lattice.

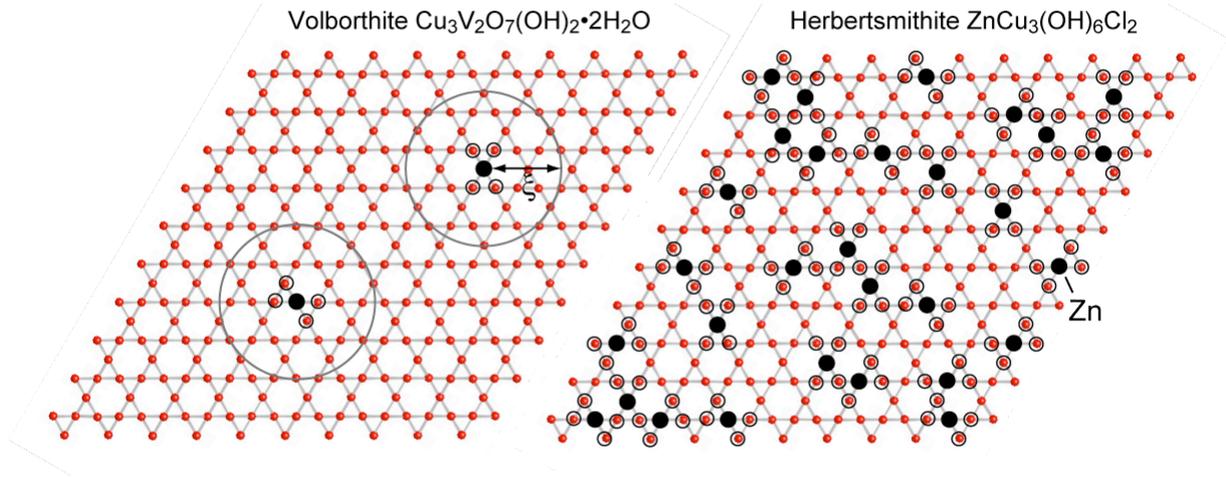

**Figure 7.** Schematic drawings of the kagome lattice for volborthite (left) and herbertsmithite (right) showing how defects are included in actual samples. Small balls represent Cu atoms, and larger closed circles some defects for volborthite and Zn atoms for herbertsmithite. It is assumed for volborthite that two defects exist among approximately 300 Cu ions, so that $x_{imp} \sim 0.7\%$, while, for herbertsmithite, Zn occupies 10% of the Cu sites randomly. Since 4 Cu spins next to a defect that are marked by open circles lose their neighbours and are affected seriously, the number of intact spins decreases rapidly with increasing $x_{imp}$. Even in a case with fewer defects, as in volborthite, the influence of defects on the background of singlet "sea" may spread over a correlation length $\xi$ that should increase with decreasing temperature until it is limited by the average separation of defects.

## 5. Concluding remarks

The GS of the spin-1/2 HAFM is still mysterious. The broad maximum and the finite value of $\chi_{bulk}$ as well as in $\chi_{local}$ observed in volborthite must represent its intrinsic properties, as it is also observed in $\chi_{local}$ for herbertsmithite. Thus, SRO develops below $T \sim J/4$, and the spin excitation at $T = 0$ may be gapless, in disagreement with theoretical predictions. The correlation length may rapidly increase with decreasing temperature, more rapidly than in one-dimensional systems, and saturate at a very large value, so that even a small proportion of defects must seriously influence their surroundings. This may be common for all the frustrated systems that lack long-range order. In the absence of defects, the GS is considered to be a sort of long-range RVB state or a spin liquid, which contains various ranges of singlet pairs over long distances. We believe that there must exist extremely slow spin dynamics originating from frustration and quantum fluctuations in the spin-1/2 KAFM, which will be addressed in future studies using a higher-quality powder sample or, hopefully, a single crystal of volborthite, or other more ideal kagome compounds.

## Acknowledgments

We thank M. Hanawa and N. Kobayashi for their contributions to our initial research into volborthite. This work was supported by Grant-in-Aid for Scientific Research on Priority Areas "Novel States of Matter Induced by Frustration" (19052003).

**References**

[1]     Misguich G and Lhuillier C, in *Furstrated Spin Systems*, edited by H. T. Diep (World Scientific, Singapole, 2005), pp. 229.
[2]     Ramirez A P 1994 *Annu. Rev. Mater. Sci.* **24** 453
[3]     Elstner N and Young A P 1994 *Phys. Rev. B* **50** 6871
[4]     Waldtmann C, Everts H-U, Bernu B, Lhuillier C, Sindzingre P, Lecheminant P and Pierre L 1998 *Eur. Phys. J. B* **2** 501
[5]     Mila F 1998 *Phy. Rev. Lett.* **81** 2356
[6]     Lafontaine M A, Bail A L and Férey G 1990 *J. Solid State Chem.* **85** 220
[7]     Hiroi Z, Kobayashi N, Hanawa M, Nohara M, Takagi H, Kato Y and Takigawa M 2001 *J. Phys. Soc. Jpn.* **70** 3377
[8]     Fukaya A, Fudamoto Y, Gat I M, Ito T, Larkin M I, Savici A T, Uemura Y J, Kyriakou P P, Luke G M, Rovers M T, Kojima K M, Keren A, Hanawa M and Hiroi Z 2003 *Phys. Rev. Lett.* **91** 207603
[9]     Okubo S, Ohta H, Hazuki K, Sakurai T, Kobayashi N and Hiroi Z 2001 *Physica B* **294-295** 75
[10]    Wang F, Vishwanath A and Kim Y B 2007 *Phys. Rev. B* **76** 094421
[11]    Sindzingre P 2008 *arXiv*:0707.4264
[12]    Yavors'kii T, Apel W and Everts H-U 2007 *Phys. Rev. B* **76** 064430
[13]    Shores M P, Nytko E A, Bartlett B M and Nocera D G 2005 *J. Am. Chem. Soc.* **127** 13462
[14]    Braithwaite R S W, Mereiter K, Paar W H and Clark A M 2004 *Mineralogical Magazine* **68** 527
[15]    Helton J S, Matan K, Shores M P, Nytko E A, Bartlett B M, Yoshida Y, Takano Y, Suslov A, Qiu Y, Chung J H, Nocera D G and Lee Y S 2007 *Phy. Rev. Lett.* **98** 107204
[16]    Rigol M and Singh R R P 2007 *Phy. Rev. Lett.* **98** 207204
[17]    Misguich G and Sindzingre P 2007 *Eur. Phys. J. B* **59** 305
[18]    Mizuno Y, Tohyama T, Maekawa S, Osafune T, Motoyama N, Eisaki H and Uchida S 1998 *Pys. Rev. B* **57** 5326
[19]    Lee S-H, Kikuchi H, Qiu Y, Lake B, Huang Q, Habicht K and Kiefer K 2007 *Nat. Mater.* **6** 853
[20]    de Vries M A, Kamenev K V, Kockelmann W A, Sanchez-Benitez J and Harrison A 2008 *Phy. Rev. Lett.* **100** 157205
[21]    Olariu A, Mendels P, Bert F, Duc F, Trombe J C, Vries M A d and Harrison A 2008 *Phys. Rev. Lett.* **100** 087202
[22]    Mendels P, Bert F, Vries M A d, Olariu A, Harrison A, Duc F, Trombe J C, Lord J S, Amato A and Baines C 2007 *Phys. Rev. Lett.* **98** 077204
[23]    Yoshida H, Okamoto Y, Tayama T, Sakakibara T, Tokunaga M, Matsuo A, Narumi Y, Kindo K, Yoshida M, Takigawa M and Hiroi Z *in preparation*
[24]    Bonner J C and Fisher M E 1964 *Phys. Rev.* **135** A640
[25]    Bert F, Bono D, Mendels P, Ladieu F, Duc F, Trombe J-C and Millet P 2005 *Phys. Rev. Lett.* **95** 087203
[26]    Takigawa M, Motoyama N, Eisaki H and Uchida S 1997 *Phys. Rev. B* **55** 14129